\documentclass[12pt,preprint]{aastex}

\usepackage{amssymb}

\newcommand{\scinum}[2]{#1 \times 10^{#2}}

\newcommand{\scinuma}[3]{(#1 \pm #2) \times 10^{#3}}

\newcommand{\decnuma}[2]{#1 \pm #2}

\newcommand{\mbh}{M_{\bullet}}

\newcommand{\msun}{\mathrm{M}_{\odot}}

\shorttitle{X-ray and UV Variability of NGC~4395} \shortauthors{P.  M.
O'Neill et~al.}

\begin{document}

\title{Multiwavelength  Monitoring  of   the  Dwarf  Seyfert~1  Galaxy
NGC~4395.  II.  X-ray and Ultraviolet Continuum Variability}

\author{Paul M. O'Neill\altaffilmark{1}, Shai Kaspi\altaffilmark{2,3},
Ari  Laor\altaffilmark{2},  Kirpal  Nandra\altaffilmark{1}, Edward  C.
Moran\altaffilmark{4},     Bradley    M.     Peterson\altaffilmark{5},
Louis-Benoit        Desroches\altaffilmark{6},        Alexei        V.
Filippenko\altaffilmark{6},  Luis   C.   Ho\altaffilmark{7},  and  Dan
Maoz\altaffilmark{3}}

\altaffiltext{1}{Astrophysics Group, Imperial College London, Blackett
Laboratory,  Prince  Consort  Road,  London SW7~2AZ,  United  Kingdom;
p.oneill@imperial.ac.uk}

\altaffiltext{2}{Department of Physics, Technion, Haifa 32000, Israel}

\altaffiltext{3}{Wise Observatory and School of Physics and Astronomy,
Tel-Aviv University, Tel-Aviv~69978, Israel}

\altaffiltext{4}{Astronomy Department,  Wesleyan University, Van Vleck
Observatory, Middletown, CT~06459}

\altaffiltext{5}{Department of  Astronomy, Ohio State  University, 140
West 18th Avenue, Columbus, OH~43210}

\altaffiltext{6}{Department  of Astronomy,  University  of California,
Berkeley, CA~94720-3411}

\altaffiltext{7}{Carnegie  Observatories,  813  Santa Barbara  Street,
Pasadena, CA~91101}

\begin{abstract}

We  report  on two  \emph{Chandra}  observations,  and a  simultaneous
\emph{Hubble  Space Telescope} ultraviolet  observation, of  the dwarf
Seyfert~1  galaxy  NGC~4395.  Each  \emph{Chandra}  observation had  a
duration  of  $\sim$30~ks,  with  a separation  of  $\sim$50~ks.   The
spectrum  was observed  to  harden between  these  observations via  a
scaling down of the soft-band flux.  The inter-observation variability
is  in a  different  sense  to the  observed  variability within  each
observation  and is most  likely the  result of  increased absorption.
Spectral variations were seen  during the first observation suggesting
that  the X-ray  emission is  produced in  more than  one disconnected
region.   We  have   also  re-analyzed  a  $\sim$17~ks  \emph{Chandra}
observation  conducted  in  2000.   During  the  three  \emph{Chandra}
observations the 2--10~keV flux is about a factor of 2 lower than seen
during an \emph{XMM-Newton}  observation conducted in 2003.  Moreover,
the   fractional   variability    amplitude   exhibited   during   the
\emph{XMM-Newton} observation is significantly softer than seen during
the  \emph{Chandra} observations.   A power-spectral  analysis  of the
first of  the two new  \emph{Chandra} observations revealed a  peak at
341~s with a  formal detection significance of 99~\%.   A similar peak
was  seen previously in  the 2000  \emph{Chandra} data.   However, the
detection  of this feature  is tentative  given that  it was  found in
neither the second of our  two new \emph{Chandra} observations nor the
\emph{XMM-Newton} data,  and it is  much narrower than  expected.  The
\emph{Hubble Space Telescope} observation was conducted during part of
the second  \emph{Chandra} visit.  A zero-lag  correlation between the
ultraviolet  and X-ray  fluxes  was detected  with  a significance  of
$\sim$99.5~\%, consistent with the  predictions of the two-phase model
for the X-ray emission from active galactic nuclei.

\end{abstract}

\keywords{galaxies:            active---galaxies:           individual
(NGC~4395)---galaxies:    Seyfert---ultraviolet:    galaxies---X-rays:
galaxies}

\section{INTRODUCTION}

The  Sd  galaxy  NGC~4395  harbors  a  Seyfert~1  nucleus  having  the
lowest-known  luminosity and  black hole  mass for  any object  of its
class  \citep[][]{fs89,fhs93,fh03}.   A  recent reverberation  mapping
campaign    yielded    a    black     hole    mass    of    $\mbh    =
\scinuma{3.6}{0.6}{5}$~$\msun$  and   a  luminosity  corresponding  to
0.1~\% of the Eddington limit \citep[][]{pbd05}.  NGC~4395 thus offers
the opportunity to study active  galactic nuclei (AGNs) at the low end
of  the mass scale.   This is  particularly useful  in the  context of
unifying  our  understanding  of  the  X-ray emission  from  AGNs  and
galactic black hole binaries.

NGC~4395 exhibits  strong and rapid X-ray variability.   The best data
in this regard are  from a recent \emph{XMM-Newton} observation, which
provided an almost uninterrupted  light curve of $\sim$90~ks duration.
The  power spectrum of  the 0.2--10~keV  variability could  be modeled
with a singly broken power  law, with a break timescale of $\sim$500~s
\citep[][]{vif04}.   Moreover,  the amplitude  of  the variations  was
found to decrease with  increasing energy.  X-ray spectral variability
in  NGC~4395 had been  studied previously  using both  \emph{ASCA} and
\emph{Chandra}  data.   The  \emph{ASCA}  data  were  modeled  with  a
multi-zone warm absorber, and  the spectral variability was suggestive
of flux-correlated  variations in the  ionization parameter of  one of
these absorbers \citep[][]{ifa00,sif03}.   In the \emph{Chandra} data,
spectral  variations  were seen  on  a  timescale  of several  hundred
seconds and these were interpreted  to be the result of column density
fluctuations   \citep[][]{mel05}.    Contrary   to   the   \emph{ASCA}
observations,  the  \emph{Chandra} data  exhibited  only a  marginally
significant   variation   in  spectral   shape   with  overall   flux.
\citet[][]{mel05}   also  reported   the  presence   of   a  transient
$\sim$400~s modulation  in the  \emph{Chandra} data, with  a detection
significance of $\gtrsim$95~\%

Simultaneous X-ray and ultraviolet (UV) observations are very valuable
in the  study of AGNs.  In  the two-phase model for  the production of
the  X-ray emission, seed  photons from  an optically  thick accretion
disk are inverse-Compton scattered by hot electrons in the disk corona
\citep[e.g.,][]{st80,hm93}.  A  portion of the X-ray  flux is directed
down toward  the disk  to be thermally  reprocessed into  seed photons
\citep[][]{gr88}. Given that the two phases (i.e. disk and corona) are
so  closely  coupled,  simultaneous   UV  and  X-ray  data  can  yield
information       regarding        the       disk--corona       system
\citep[e.g.,][]{nlg00,pmh04}.

In this  paper, we present  a variability analysis of  two $\sim$30~ks
\emph{Chandra}  observations   that  were   obtained  as  part   of  a
multiwavelength monitoring campaign on  NGC~4395. The results from the
UV  line variability  analysis  from this  campaign  are presented  in
\citet[][]{pbd05}, and the optical  results are described by Desroches
et~al.   (2006).  We  have also  re-analyzed the  previous $\sim$17~ks
\emph{Chandra}  observation  from  2000  and,  for the  purpose  of  a
comparison  with  the \emph{Chandra}  data,  we  have re-analyzed  the
$\sim$90~ks   \emph{XMM-Newton}   observation.    Finally,   we   have
investigated  the X-ray--UV relationship  using the  simultaneous data
obtained during the second \emph{Chandra} observation.

%%%%%%%%%%%%%%%%%%%%%%%%%%%%%%%%%%%%%%%%%%%%%%%%%%%%%%%%%%%%%%%%%%%%%%%%
%%%%%%%%%%%%%%%%%%%%%%%%%%%%%%%%%%%%%%%%%%%%%%%%%%%%%%%%%%%%%%%%%%%%%%%%%

\section{OBSERVATIONS AND DATA REDUCTION}

\emph{Chandra}  observed NGC~4395  on  2004 April  10,  from 02:43  to
11:39~UT, and on April 11, from 01:05 to 10:11~UT.  We hereafter refer
to the  first and second observations as  ``visit~1'' and ``visit~2,''
respectively. The observations were conducted using the S3 chip of the
ACIS-S instrument.   A 1/8 subarray was selected,  with each 0.44104~s
frame  having  an  actual  exposure  time of  0.4~s  (i.e.,  live-time
$\sim$91~\%).     The   \emph{Chandra}   observation    presented   by
\citet[][see \S~1]{mel05} was conducted on 2000 June 20, from 03:04 to
07:56~UT, and we  shall refer to it by  its observation identification
number 882.   A 1/2 subarray  was selected for observation  882, which
yielded frame and exposure  times of 1.54104~s and 1.5~s, respectively
(i.e. live-time  $\sim$97~\%).  Level~2  events files were  created in
the   standard   manner  using   CIAO~3.2.1   and  CALDB~3.0.1.    All
observations were uninterrupted and no background flares were present.
The  total  good exposure  times  were  $\sim$27~ks, $\sim$28~ks,  and
$\sim$17~ks for  visit~1, visit~2, and  observation~882, respectively.
\citet[][]{mel05}  showed that photon  pile-up does  not significantly
affect observation~882,  and the shorter frame time  used for visits~1
and 2  ensures that  pile-up is not  a problem for  those observations
either.

Source events were extracted from within a circular aperture, centered
on  the source,  with  a radius  of  10~pixels ($\sim$5\arcsec).   The
background  was  estimated  from  an annulus  surrounding  the  source
region, with inner and outer  radii of 20 and 55 pixels, respectively.
Light curves  were then  extracted from these  events files,  with the
time resolution  chosen to be an  integer multiple of  the frame time.
For visit~1 and visit~2, we  used time bins of 299.9072~s, 599.8144~s,
and  999.83768~s,  respectively,  for  the ``300~s,''  ``600~s,''  and
``1000~s'' light  curves.  For observation~882,  we used time  bins of
300.5028~s, 601.0056~s, and 1000.13496~s.  The data were resolved into
three  (0.4--2.7,  2.7--4.2,  4.2--8.0~keV)  and also  two  (0.4--3.4,
3.4--8.0~keV) energy  ranges.  The  boundaries of these  energy ranges
were selected so  that each band has roughly  the same signal-to-noise
ratio.   We  shall  refer   to  the  0.4--8.0~keV,  0.4--3.4~keV,  and
3.4--8.0~keV  energy ranges  as the  ``full,'' ``soft,''  and ``hard''
bands, respectively.   Furthermore, we  define ``softness'' to  be the
ratio  between  the counting  rates  in  these  soft and  hard  bands.
Resolving the data into three energy  bands, with a time bin of 600~s,
yields an average of roughly 20~counts~bin$^{-1}$.

As well as binning the  events in equal time intervals, we constructed
light curves with a variable bin width. The start time of each bin was
defined  to be  the midpoint  in time  between two  consecutive photon
arrival times. The length of each time bin was then set such that each
bin contained exactly 30~counts.  The stop time of the bin was defined
to be the midpoint in time between  the last count in that bin and the
next event.   Smoothing was achieved  by using overlapping  bins, such
that adjacent bins in the smoothed  light curves share a common set of
29 counts.

\begin{figure}
\epsscale{0.8} \plotone{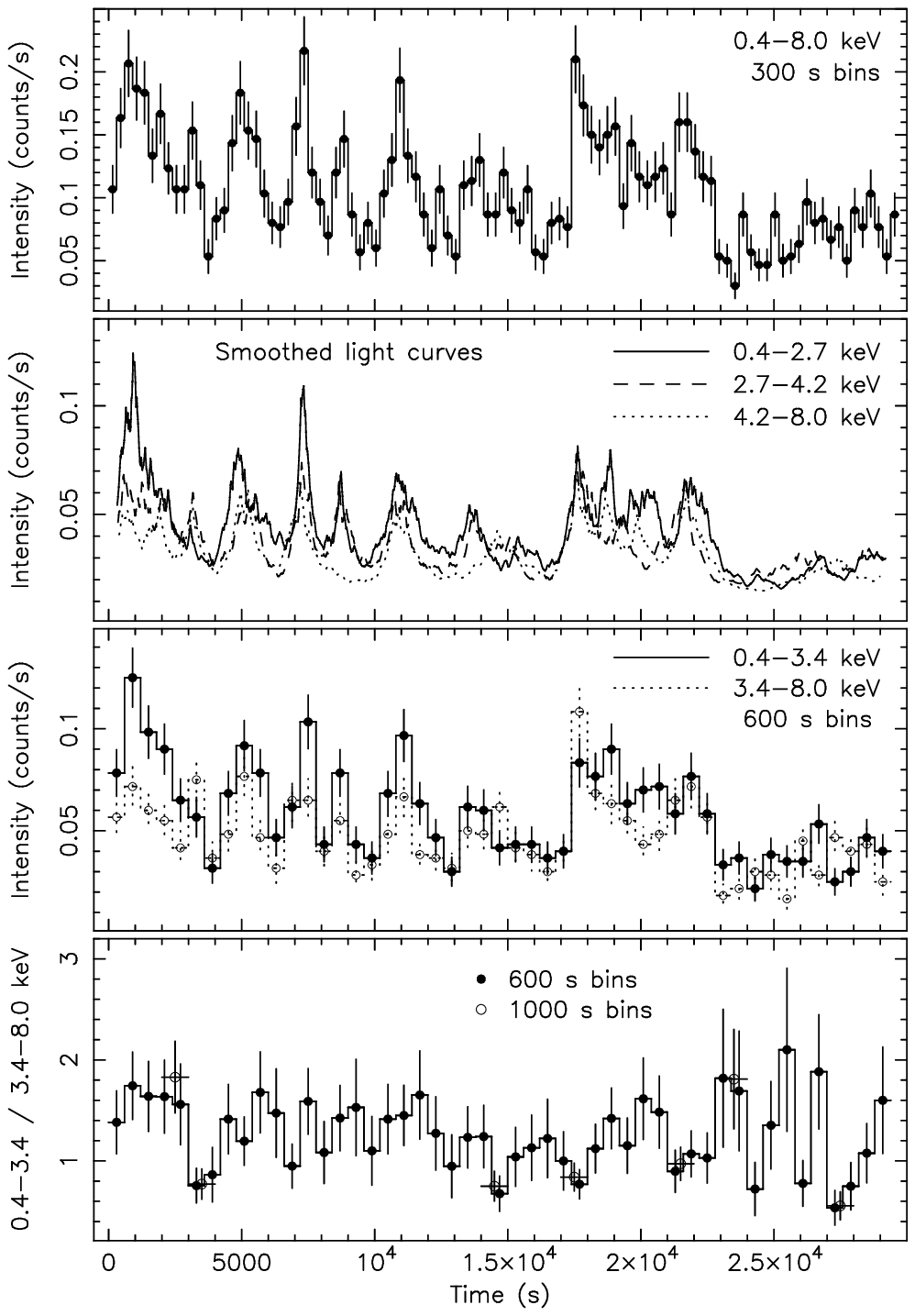}

\caption{Full-band (0.4--8.0~keV) and energy-resolved light curves for
visit~1.  The smoothed light  curves show the running means calculated
from overlapping groups of 30 consecutive events. In the bottom panel,
the solid  circles show the  softness calculated from 600~s  bins, and
the open  circles correspond to  1000~s bins.  These open  circles are
shown for the  seven points that each contribute  $\Delta\chi^{2} > 2$
to the straight-line  fit between the soft and  hard intensities shown
in Fig.~\ref{fig:visit1_2_882_softvhard}.\label{fig:visit1_all}}

\end{figure}

\begin{figure}
\epsscale{0.8} \plotone{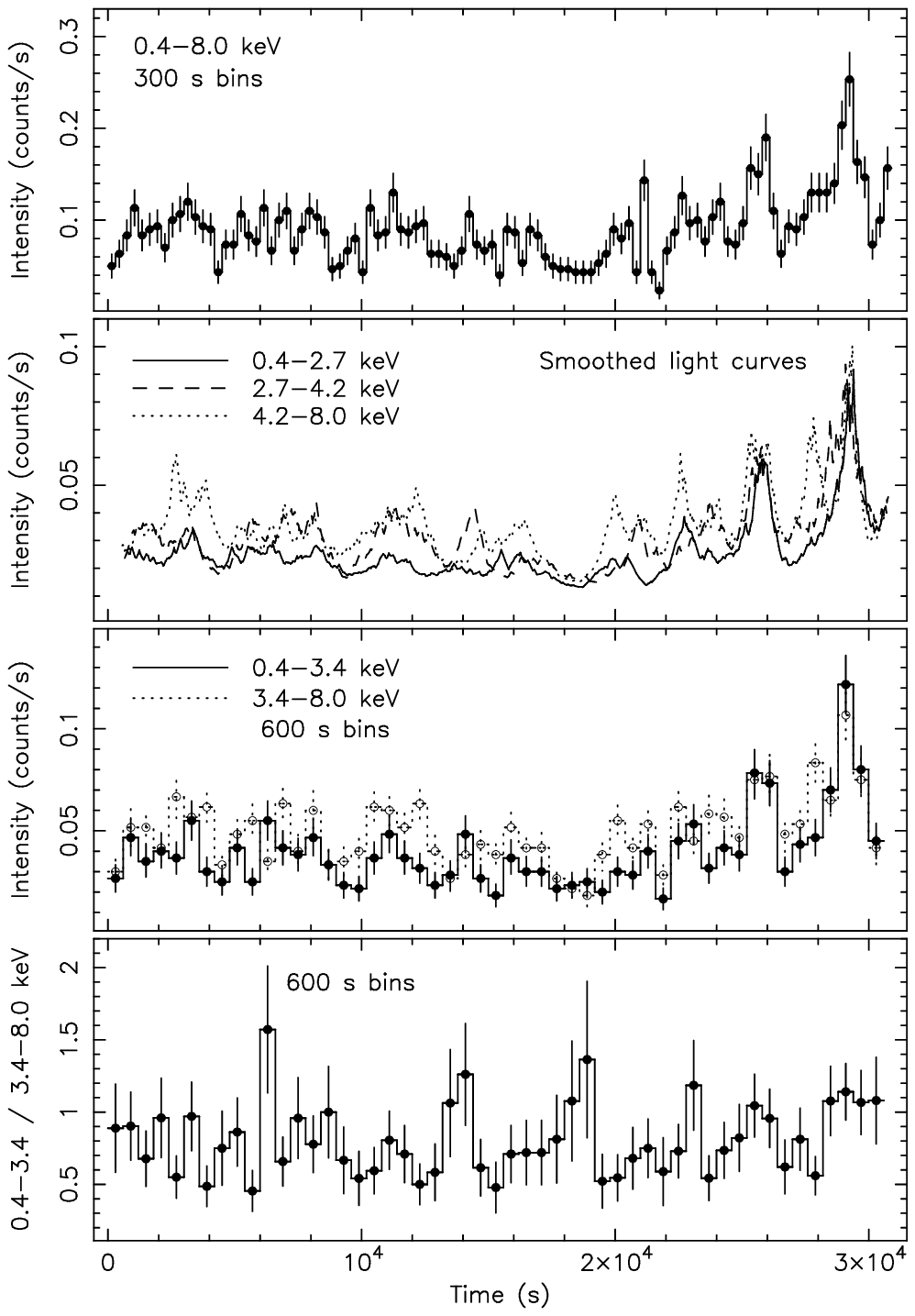}

\caption{Full-band (0.4--8.0~keV) and energy-resolved light curves for
visit~2. The  smoothed light curves show the  running means calculated
from       overlapping       groups       of      30       consecutive
events.\label{fig:visit2_all}}

\end{figure}

\begin{figure}
\epsscale{0.8} \plotone{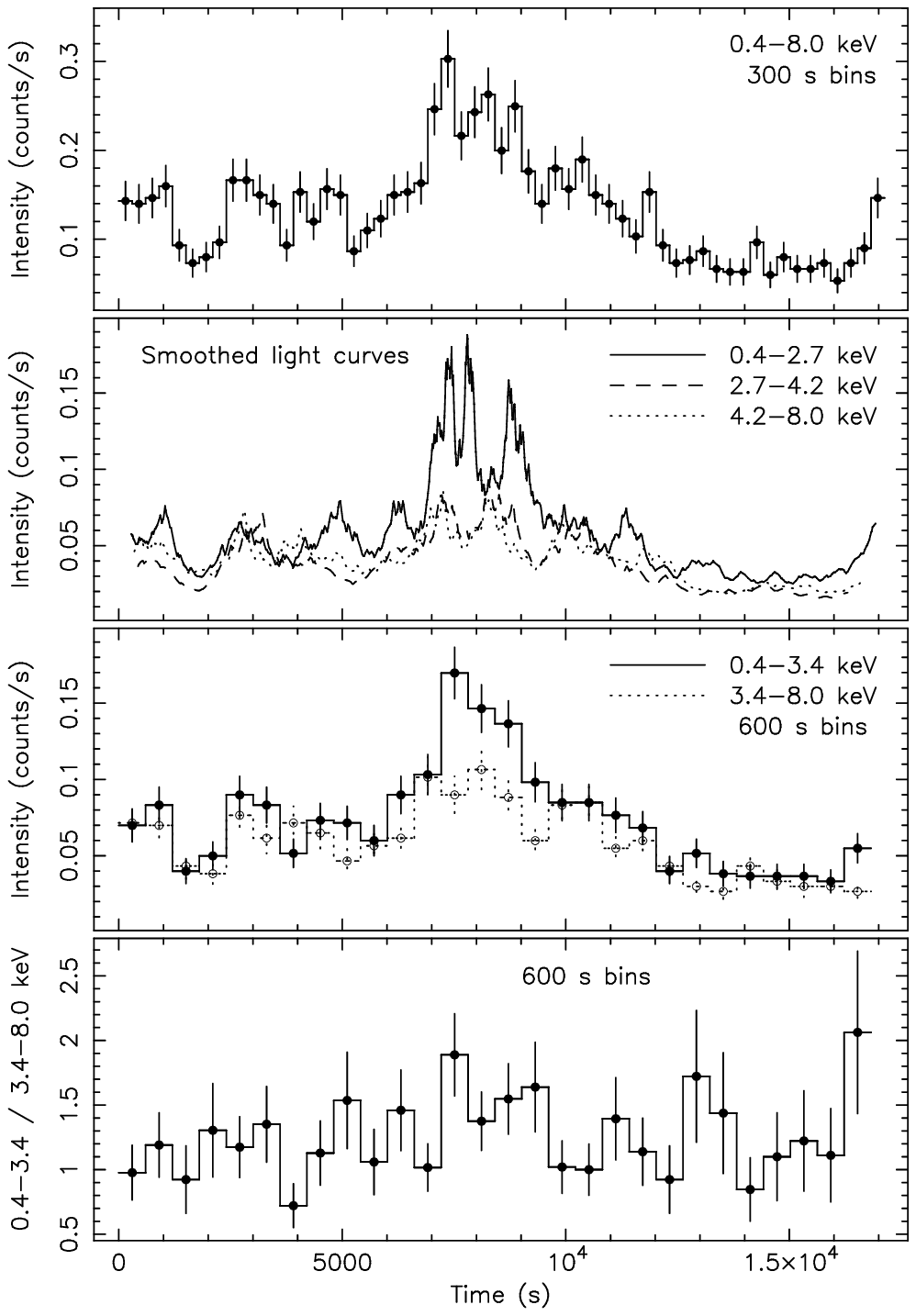}

\caption{Full-band (0.4--8.0~keV) and energy-resolved light curves for
observation  882.  The smoothed  light curves  show the  running means
calculated    from    overlapping    groups    of    30    consecutive
events. \label{fig:obs882_all}}

\end{figure}

Various  light curves for  visit~1, visit~2,  and observation  882 are
presented  in  Figs.~\ref{fig:visit1_all},  \ref{fig:visit2_all},  and
\ref{fig:obs882_all}, respectively.   The top  panel in each  of these
figures    shows   the    full-band   (0.4--8.0~keV),    300~s   light
curve. Smoothed  light curves, resolved  into three energy  bands, are
shown in the second panel. In  the third panel we show 600~s soft-band
and hard-band light curves. Finally,  the softness ratio using a 600~s
time  resolution is  given  in the  lowest  panel.  In  this panel  in
Fig.~\ref{fig:visit1_all}, we have overlaid 7 points from the softness
time  series  using 1000~s  bins.   The  meaning  of these  points  is
explained in  \S~\ref{sec:cor}. The error bars are  1$\sigma$, and the
time origins are MJD~53105.12824, MJD~53106.06050, and MJD~51715.13164
for visit~1, visit~2, and observation~882, respectively.

The visit~1  and visit~2 {\it  Chandra} observations were  carried out
simultaneously  with {\it  HST} Space  Telescope  Imaging Spectrograph
(STIS)  observations, which  are described  by  \citet[][]{pbd05}. The
\emph{HST} observation during visit~1 was acquired under gyro control,
which degraded the photometric  accuracy and rendered this observation
unusable.  Therefore, our X-ray--UV correlation analysis uses only the
visit~2 data. Note that, owing  to a data processing error, the fluxes
given by  \citet[][]{pbd05} are too high  by a factor of  7.96. The UV
fluxes presented here have been scaled to the correct values.

We have  also performed an analysis of  the archived \emph{XMM-Newton}
data on  NGC~4395 to facilitate  a comparison with  the \emph{Chandra}
data.  This observation was conducted on 2003 November 30 and December
1  for  $\sim$110~ks.   We  used   only  the  earliest  90~ks  of  the
observation,  during  which the  background  was  low  and stable.   A
variability analysis of these data was presented by \citet[][]{vif04}.

%%%%%%%%%%%%%%%%%%%%%%%%%%%%%%%%%%%%%%%%%%%%%%%%%%%%%%%%%%%%%%%%%%%%%%%%
%%%%%%%%%%%%%%%%%%%%%%%%%%%%%%%%%%%%%%%%%%%%%%%%%%%%%%%%%%%%%%%%%%%%%%%%

\section{X-RAY VARIABILITY}

\subsection{Normalized Excess Variance}

One  method of investigating  spectral variability  is to  compute the
excess  variance in  the  light curve  as  a function  of energy.   We
sub-divided  the   visit~1  and  visit~2  light   curves  (600~s  time
resolution)  each  into  2  halves,  thus creating  four  light  curve
segments having  roughly the same  duration as the entire  light curve
from  observation~882.  We  resolved these  light curves  into various
energy bands and measured  the normalized excess variance \citep[i.e.,
excess variance divided by the mean-squared; see, e.g.,][]{vew03}.

The  uncertainties were calculated  directly from  the data  via Monte
Carlo  simulations.    For  each   light  curve,  we   performed  1000
simulations  in which  we added  Poisson noise  to the  observed light
curve  and calculated  the normalized  excess variance.   The standard
deviation of  the 1000 simulated values of  normalized excess variance
was  then taken  to be  the uncertainty  owing \emph{only}  to Poisson
noise. For  each energy band, we  determined the mean of  each pair of
measurements  from visit~1  and  visit~2, and  the uncertainties  were
propagated.

The  values of normalized  excess variance  for visit~1,  visit~2, and
observation~882 are listed in Table~\ref{tab:snxs}, where we also give
the  mean variance  obtained  from all  5  \emph{Chandra} light  curve
segments. The ratios between  the excess variances in the 0.4--2.7~keV
and 4.2--8.0~keV  bands are $1.5\pm0.6$,  $1.5\pm0.7$, and $2.4\pm0.9$
for visit~1, visit~2, and observation~882, respectively.  These ratios
are  all consistent  with unity,  and are  thus consistent  with there
being  no  spectral variability  during  each  of these  observations.
Considering all of the \emph{Chandra} light curves combined, the ratio
between  the excess  variances  in the  2.7--4.2~keV and  4.2--8.0~keV
bands  is $1.4\pm0.3$, which  is also  clearly consistent  with unity.
However, the ratio between  the 0.4--2.7~keV and 4.2--8.0~keV bands is
$1.7\pm0.4$,  providing marginal  ($\sim$2$\sigma$)  evidence for  the
presence of enhanced soft variability.  Note also that there \emph{is}
evidence  for enhanced  soft variability  during  observation~882: the
\emph{difference} between the  0.4--2.7~keV and 4.2--8.0~keV variances
is    $\scinuma{13}{5}{-2}$,    which    is   significant    at    the
$\sim$2.5$\sigma$ level.

\begin{figure}

\plotone{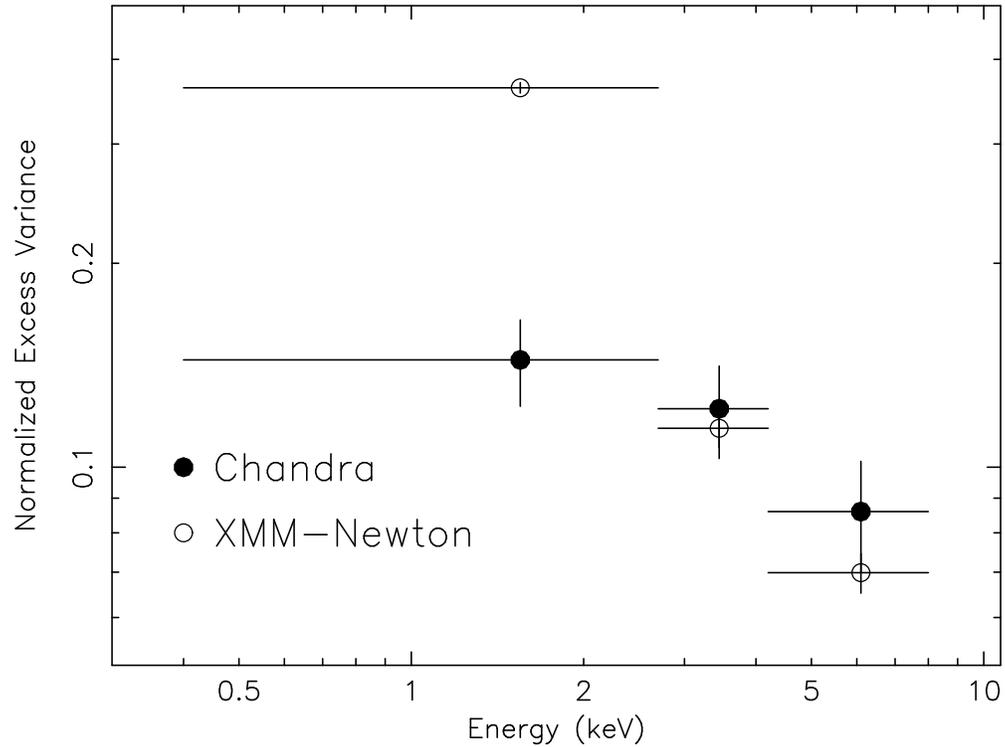}

\caption{Normalized   excess    variances   in   the    energy   bands
  0.4--2.7--4.2--8.0~keV,  for  600--15000~s  timescales.  The  filled
  circles show  the mean  variance of the  \emph{Chandra} observations
  while the open circles correspond to the \emph{XMM-Newton} data. The
  error  bars  show  the  uncertainties expected  from  Poisson  noise
  only.\label{fig:rmsspec}}

\end{figure}

We  also wished to  compare the  \emph{Chandra} observations  with the
\emph{XMM-Newton}  observation.  We  subdivided  the \emph{XMM-Newton}
light curve into 6 segments,  each with a duration of $\sim$15~ks, and
selected the  same energy  bands that we  used for  the \emph{Chandra}
data.  The  mean excess variances from the  \emph{XMM-Newton} data are
listed  in  Table~\ref{tab:snxs},  and  the variances  from  both  the
\emph{Chandra}  and  \emph{XMM-Newton}  observations  are  plotted  in
Fig.~\ref{fig:rmsspec}.   The  ratio  between  the  variances  in  the
2.7--4.2~keV   and  4.2--8.0~keV  bands   is  $1.6\pm0.2$,   which  is
consistent   with   that   seen   during  the   three   \emph{Chandra}
observations.  Considering  instead the 0.4--2.7~keV  and 4.2--8.0~keV
bands,  the ratio  between the  excess variances  is  $5.2\pm0.4$. The
variability  is  thus  clearly  softer  during  the  \emph{XMM-Newton}
observation compared to the \emph{Chandra} observations.

\subsection{Correlation Analysis} \label{sec:cor}

As  well as  examining the  relative magnitudes  of the  variations in
different  energy bands,  we have  also investigated  the correlations
that exist  between them.  This is useful  because the excess-variance
analysis is not sensitive to the presence of phase differences between
the variations in the different energy bands.

The ``softness'' was defined to be the counting rate ratio between the
0.4--3.4~keV         and        3.4--8.0~keV         bands        (see
Figs.~\ref{fig:visit1_all}--\ref{fig:obs882_all}).  During visit~2 and
observation~882, using a time resolution of both 600~s and 1000~s, the
softness is consistent with  being constant.  During visit~1, however,
the  softness is  variable with  a  significance of  97~\% and  99~\%,
respectively, when using time resolutions of 600~s and 1000~s.

\begin{figure}

\plotone{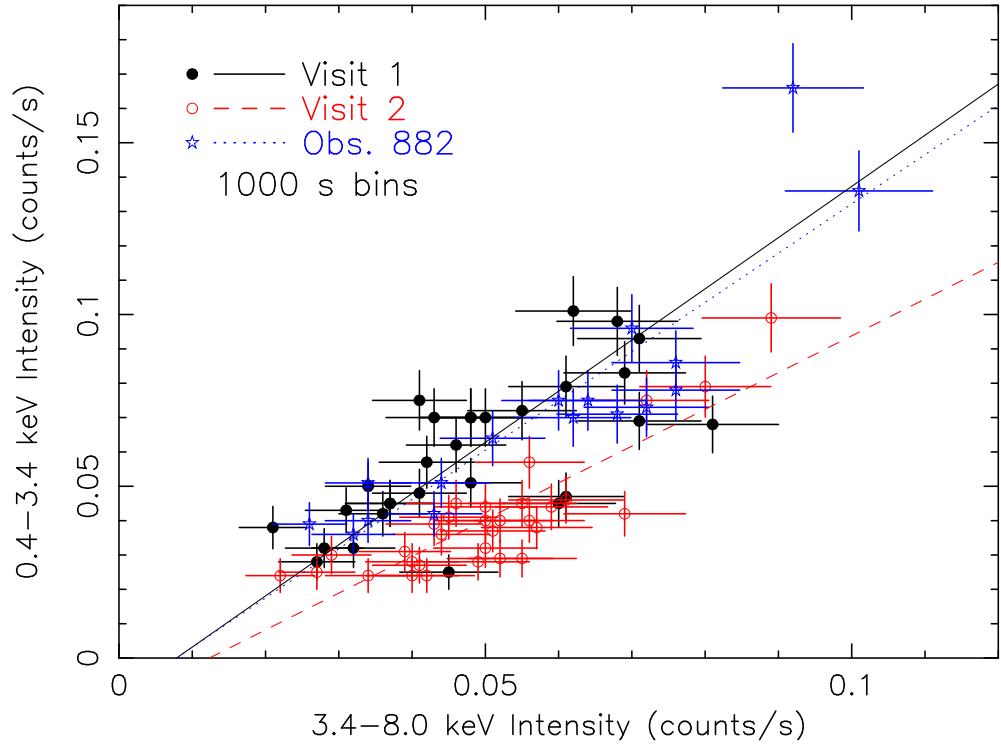}

\caption{Correlations   between  the  0.4--3.4~keV   and  3.4--8.0~keV
intensities  during  visit~1   (solid  circles,  black  in  electronic
edition),  visit~2  (open  circles,  red in  electronic  edition)  and
\emph{Chandra}  observation 882 (stars,  blue in  electronic edition).
Each  datum corresponds  to a  time bin  of 1000~s.   The best-fitting
straight lines for  visit~1, visit~2, and observation 882  are shown as
solid  (black  in  electronic  edition),  dashed  (red  in  electronic
edition),   and   dotted    (blue   in   electronic   edition)   lines,
respectively. See  the electronic edition  of the Journal for  a color
version of this figure.\label{fig:visit1_2_882_softvhard}}

\end{figure}

In  Fig.~\ref{fig:visit1_2_882_softvhard}  we  show  the  relationship
between the  0.4--3.4~keV and  3.4--8.0~keV intensities for  all three
observations,  using a  time  resolution of  1000~s.   We performed  a
chi-squared  straight-line  fit  for  each  observation,  taking  into
account  the  uncertainties  in  both  coordinates  \citep[][]{ptv01}.
These fits were satisfactory  for visit~2 and observation~882. The fit
for  visit~1, however, was  poor ($\chi^{2}/\mathrm{DOF}  = 45.0/27$),
revealing  the presence  of scatter  about the  mean relation,  with a
significance     of     98.4~\%.      The    mean     relations     in
Fig.~\ref{fig:visit1_2_882_softvhard} are  all consistent with passing
through  the origin,  so  there  is no  significant  softening of  the
spectrum with increasing flux.

In  Fig.~\ref{fig:visit1_2_882_softvhard},   there  are  seven  points
during  visit~1 that  deviate from  the straight  line such  that they
contribute $\Delta\chi^{2} > 2$  to the total $\chi^{2}$. These points
are   plotted    as   open   circles   in   the    bottom   panel   of
Fig.~\ref{fig:visit1_all}.   These   deviations  can  be  investigated
further by  resolving the  data into narrower  energy bands.   We thus
defined  hard color  to  be  the ratio  between  the 2.7--4.2~keV  and
4.2--8.0~keV intensities,  while soft color  is the ratio  between the
0.4--2.7~keV  and 4.2--8.0~keV  intensities.  We  perfomed chi-squared
tests and found that both the hard and soft colors are consistent with
being  constant over  the duration  of visit~1,  and  the relationship
between  these  colors is  consistent  with  Poisson  noise.  The  low
signal-to-noise ratio of  the data thus limits our  ability to examine
the spectral deviations in detail.

Between  visit~1  and visit~2  the  gradient  of  the linear  relation
between the 0.4--2.7~keV and  4.2--8.0~keV counting rates (1000~s time
resolution)     exhibits      a     significant     decrease,     from
$\decnuma{1.62}{0.23}$ during visit~1 to $\decnuma{0.87}{0.15}$ during
visit~2.   Between these observations,  then, the  soft-band intensity
has been  scaled down relative  to that in  the hard band.   Note also
that,  in  Fig.~\ref{fig:visit1_2_882_softvhard},  the  ``track''  for
observation~882 extends  to higher values  of hard and  soft intensity
compared to the other two observations.

Cross-correlation  functions (CCFs)  were calculated  for each  of the
\emph{Chandra} observations using a  time resolution of 600~s. We used
the  methods  described  in   \S~\ref{sec:uv}  and  the  analysis  was
performed on two pairs of energy bands: 0.4--3.4~keV and 3.4--8.0~keV,
and 0.4--2.7~keV  and 4.2--8.0~keV.   No significant lags  were found,
nor did  we find any clear  asymmetries in the CCFs.   The 95~\% upper
limits on the time lags were roughly $\pm$1~hour.

\subsection{Power-Spectral Analysis}

\begin{figure}

\plotone{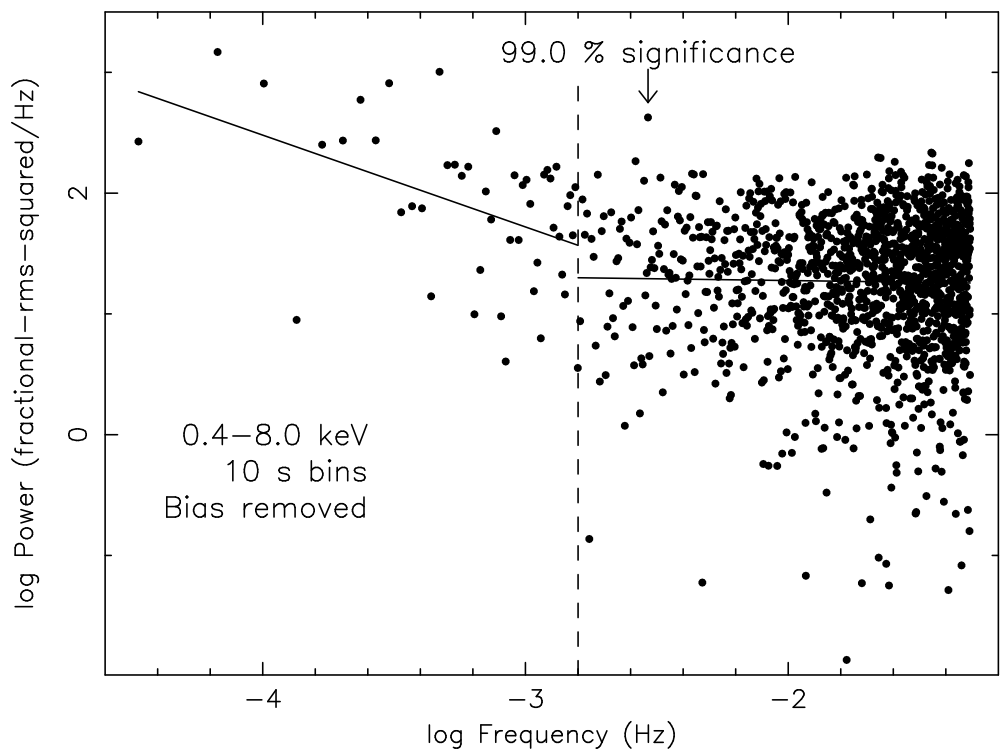}

\caption{Full-band (0.4--8.0~keV) power spectrum from visit~1, using a
time resolution of 10.14392~s. The logarithm (base 10) of the power is
plotted versus the  logarithm of frequency, and the  constant bias has
been removed (see text). The solid lines show the least-squares linear
models for frequencies below 10$^{-2.8}$~Hz (indicated with the dashed
line) and  above 10$^{-2.8}$~Hz. The most significant  power above the
continuum  is  at  a   frequency  of  $\scinum{2.93}{-3}$~Hz,  and  is
indicated with the arrow.\label{fig:visit1_pow}}

\end{figure}

A previous analysis of  observation~882 revealed some evidence for the
presence of  quasi-periodic oscillations with a  period of $\sim$400~s
\citep[][]{mel05}.  We  decided, therefore, to search  the visit~1 and
visit~2 data  for possible  periodicities. We generated  power spectra
from  the  0.4--8.0~keV  light   curves  with  a  time  resolution  of
10.14392~s  (corresponding  to  23~frames).   The frequency  step  was
$\scinum{3.36}{-5}$~Hz   in    the   visit~1   power    spectrum   and
$\scinum{3.21}{-5}$~Hz for  visit~2.  The base-10  logarithms of power
and frequency  were calculated,  and we added  0.25068 to each  of the
logarithmic powers  to remove  the bias that  is associated  with this
transformation \citep[][]{pl93,v05}.  A plot of the logarithm of power
(fractional-rms-squared~per~Hz)  versus  logarithm  of frequency  from
visit~1 is shown in Fig~\ref{fig:visit1_pow}.

We followed the prescription of \citet[][]{v05} to search for possible
periodicities.  For  visit~1, the power spectrum was  divided into two
frequency ranges,  above and  below 10$^{-2.8}$~Hz (the  Poisson noise
appears  to dominate  the power  spectrum at  frequencies  higher than
this), so that the power-law  continuum could be modeled adequately in
logarithm-logarithm    space   using    a   linear    function.    The
lower-frequency  part of  the visit~1  power spectrum  had a  slope of
$-0.8\pm0.2$, and the higher-frequency  part was consistent with white
noise.  The power  was calculated at a total  of 1464 frequencies over
the   entire   frequency   range   so   we   adopted   this   as   the
``number-of-trials.''  The most significant  peak in visit~1 was found
at a frequency of $\scinum{2.93}{-3}$~Hz, corresponding to a period of
341~s, with a significance of 99.0~\% (see Fig.~\ref{fig:visit1_pow}).
There is  no evidence for any  signal power in  the adjacent frequency
bins, so any  periodicity, if actually present, would  probably have a
quality  factor  ($Q  =  \nu/FWHM$)  of greater  than  $\sim$90.   The
detection  significance  of the  peak  might  depend  somewhat on  the
frequency at which the spectrum  is divided into two halves.  Assuming
the  division  to be  located  at  10$^{-2.5}$~Hz and  10$^{-2.45}$~Hz
(i.e.,  the peak is  on top  of the  flicker noise  component) yielded
significance  levels of  99.3~\% and  98.5~\%, respectively.   We also
examined visit~2  and the \emph{XMM-Newton} data in  a similar fashion
to visit~1 and found no evidence for a periodicity.

We  also re-analyzed observation~882  to compare  the methods  we have
used here  with the analysis  of \citet[][]{mel05}.  A  power spectrum
was  generated  from  the   0.4--8.0~keV  light  curve,  with  a  time
resolution of  9.24624~s (corresponding  to 6~frames).  In  this power
spectrum, the  boundary between the upper-  and lower-frequency halves
was chosen to  be 10$^{-2.4}$~Hz.  The most significant  peak was at a
frequency  of  $\scinum{2.50}{-3}$~Hz, corresponding  to  a period  of
400~s, with a significance of  28~\%.  Then, following the analysis of
\citet[][]{mel05}, we  generated a power spectrum using  a light curve
from only the second half of the observation.  The significance of the
peak in  this case was  95.0~\%, in agreement  with \citet[][]{mel05},
and the frequency of the peak corresponds to a period of 390~s.  (Note
that  \citet[][]{mel05} interpolated  between Fourier  frequencies and
located the peak at 396~s.)

%%%%%%%%%%%%%%%%%%%%%%%%%%%%%%%%%%%%%%%%%%%%%%%%%%%%%%%%%%%%%%%%%%%%%%%%%%%%%
%%%%%%%%%%%%%%%%%%%%%%%%%%%%%%%%%%%%%%%%%%%%%%%%%%%%%%%%%%%%%%%%%%%%%%%%%%%%%

\section{X-RAY--UV CORRELATION} \label{sec:uv}

\subsection{Cross-Correlation Analysis}

\begin{figure}

\plotone{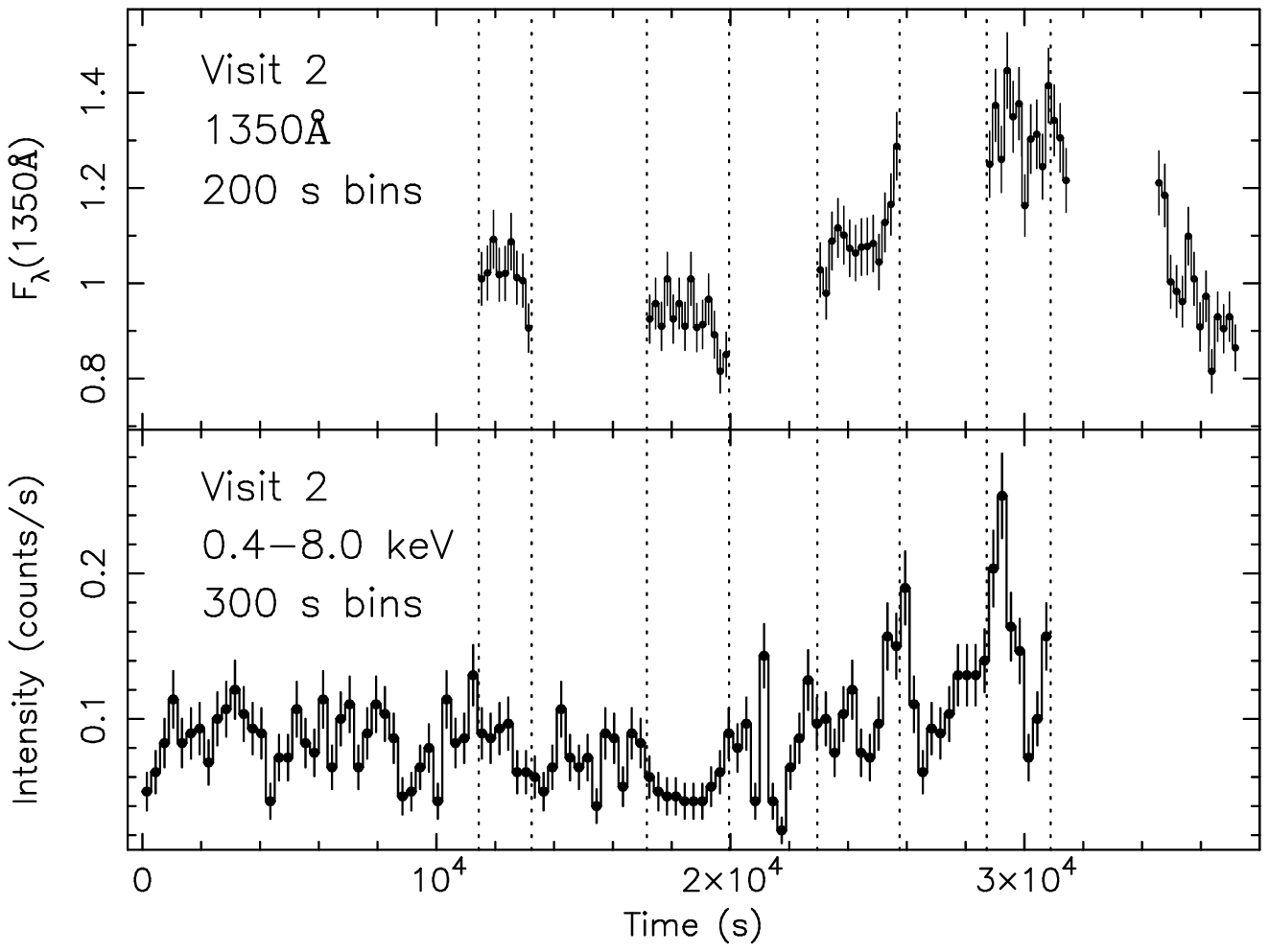}

\caption{UV continuum  flux density at 1350~\r{A}  (top) and full-band
  (0.4-8.0~keV) intensity  (bottom) during visit~2.  The  UV and X-ray
  light curves  use time resolutions of 200~s  and 300~s, respectively.
  The UV data  are from \citet[][]{pbd05} and the flux  is in units of
  10$^{-15}$~erg~cm$^{-2}$~s$^{-1}$~\r{A}$^{-1}$.    The  dotted  lines
  indicate the  boundaries of  the time intervals  having simultaneous
  X-ray and UV observations.\label{fig:uv_5302}}

\end{figure}

\begin{figure}

\plotone{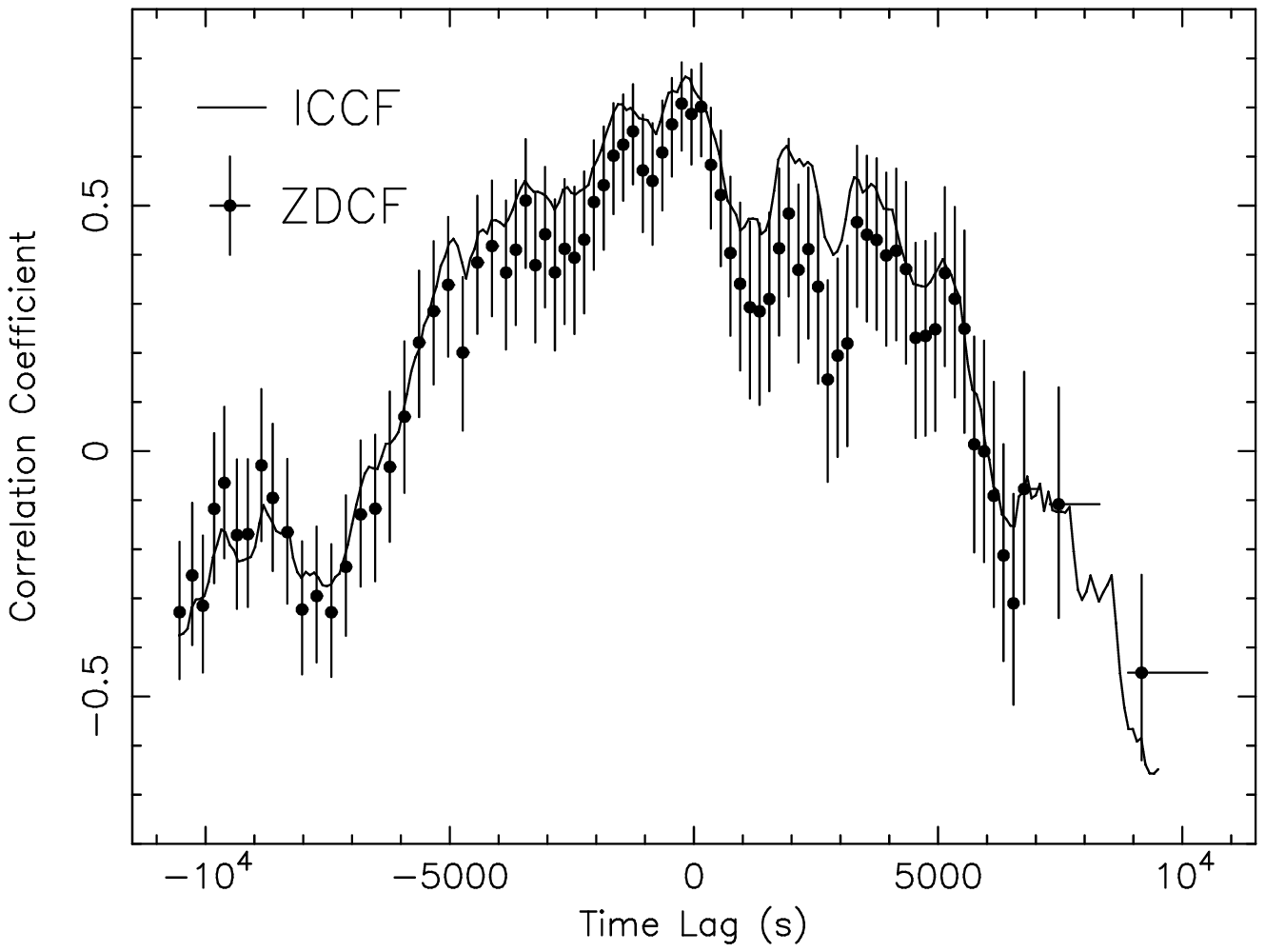}

\caption{Cross-correlation  function  between  the UV  continuum  flux
density at  1350~\r{A} and the  0.4--8.0~keV intensity. The  solid line
and  solid symbols  correspond,  respectively, to  the  ICCF and  ZDCF
methods.\label{fig:uvx_ccf}}

\end{figure}

The {\it HST/STIS} light curve at  1350 \AA, with a time resolution of
200~s,   is   shown    in   Fig.~\ref{fig:uv_5302}~(top).    We   have
cross-correlated this  light curve  with the 0.4--8.0~keV  light curve
binned  to 300~s, shown  in Fig.~\ref{fig:uv_5302}~(bottom).   We used
two   cross-correlation   methods,    one   being   the   interpolated
cross-correlation function \citep[ICCF; e.g.,][]{gs86,wp94,pfg04}, and
the other  being the z-transform discrete  correlation function (ZDCF)
of  \citet[][]{a97},   which  is   an  improvement  of   the  discrete
correlation  Function   \citep[DCF;][]{ek88}.   The  results   of  the
cross-correlations are plotted  in Fig.~\ref{fig:uvx_ccf}, in which we
show lags that  are symmetrical about zero and for  which there are at
least 20~lags for each bin in the ZDCF.  The two different methods are
in very good agreement for these data.

To   calculate  the   uncertainties  in   the   cross-correlation  lag
determination we used the  model-independent FR/RSS Monte Carlo method
of  \citet[][]{pbd05,pwh98}.    In  this  method,   each  Monte  Carlo
simulation is composed  of two parts.  The first  is a ``random subset
selection'' (RSS)  procedure which consists of  randomly drawing, with
replacement, from  a light  curve of  $N$ points a  new sample  of $N$
points. After  the $N$ points  are selected, the  redundant selections
are  removed from  the  sample such  that  the temporal  order of  the
remaining points  is preserved. This  procedure reduces the  number of
points in each light curve by  a factor of $\sim 1/e$ and accounts for
the effect that individual  data points have on the cross-correlation.
The second part is ``flux  randomization'' (FR), in which the observed
fluxes  are  altered  by   random  Gaussian  deviates  scaled  to  the
uncertainty  ascribed to  each  point.  This  procedure simulates  the
effect of  measurement uncertainties.   The two resampled  and altered
time  series  are  cross-correlated  using  the ICCF  method  and  the
centroid  of the CCF  is computed.   We used  $\sim 1000$  Monte Carlo
realizations  to build  up a  cross-correlation  centroid distribution
\citep[CCCD;][]{mz89}. The mean of the distribution is taken to be the
time lag and the uncertainty  is determined as the range that contains
68~\%  of the  Monte Carlo  realizations in  the CCCD  and  thus would
correspond to  1$\sigma$ uncertainties for a  normal distribution.  We
find  the  time lag  between  the  UV and  X-ray  light  curves to  be
$\decnuma{-300}{1400}$~s, which  is clearly consistent  with zero time
lag, with a 95~\% upper limit of roughly $\pm$1~hour.

\subsection{Light-Curve Simulations}

While  the  X-ray--UV  correlation   appears  to  be  significant,  an
apparently  strong  correlation  could  be  the  result  of  a  chance
alignment of two uncorrelated ``flicker-noise'' or ``red-noise'' light
curves.   If the  power  spectrum  of a  light  curve exhibits  either
flicker-noise  or red-noise,  then  a correlation  will exist  between
points that  are nearby  in time.  Consequently,  the \emph{effective}
number  of independent points  in each  light curve  is less  than the
actual number  of points.  Therefore, in  the case that  the two light
curves  are  uncorrelated,  the  probability of  exceeding  a  certain
correlation coefficient is  greater than in the case  of comparing two
light curves that originate from a ``white-noise'' process.

To  test  whether  we  have  observed a  significant  correlation,  we
simulated  X-ray light  curves and  cross-correlated them  against the
observed  UV   light  curve.   The  power  spectrum   of  the  visit~2
0.4--8.0~keV  light curve  (see  Fig.~\ref{fig:uv_5302}~bottom) had  a
power-law index of $\alpha =  0.86\pm0.12$, and we used this value for
the underlying mean  power spectrum.  We initially used  the method of
\citet[][]{tk95} to simulate linear light curves.  However, it was not
possible to reproduce the observed fractional variance unless negative
counting     rates    were     permitted.      Therefore,    following
\citet[][]{umv05}, we  performed an exponential  transformation on the
linear light curves. We were thus able to simulate light curves having
the  required power-law  index  and variance,  while also  maintaining
positive   counting  rates.   As   shown  by   \citet[][]{umv05},  the
exponential  transformation slightly  alters  the slope  of the  power
spectrum.  We  adopted a method  of trial-and-error to  generate power
spectra having  the required mean  power-law index. Note also  that an
exponential transform  is more appropriate  than using a  linear model
because   the   transformed  light   curves   contain  the   so-called
``rms-flux''  correlation,   which  has  been   observed  in  NGC~4395
\citep[][]{vif04}.

For  each synthesized  X-ray light  curve we  calculated the  ZDCF and
found the largest  correlation coefficient in the range  of lags shown
in  Fig.~\ref{fig:uvx_ccf}.   We  found  that  only 50  of  the  10000
simulations  had  a  coefficient  larger  than  that  observed,  which
corresponds to  a detection significance  of 99.5~\%.  It  is possible
that the true, underlying power  spectrum has a steeper index than the
best-fitting  value  of  0.86,  so  this  significance  level  may  be
overestimated.   If  we  adopt  instead  a value  of  1.10,  which  is
2$\sigma$ higher than the best-fitting value, then the significance is
reduced to  99.4~\%.  Alternatively, the significance  is increased to
99.8~\% if we use a flatter slope of $\alpha = 0.65$.

\subsection{Simultaneous UV--X-ray Observations}

\begin{figure}

\plotone{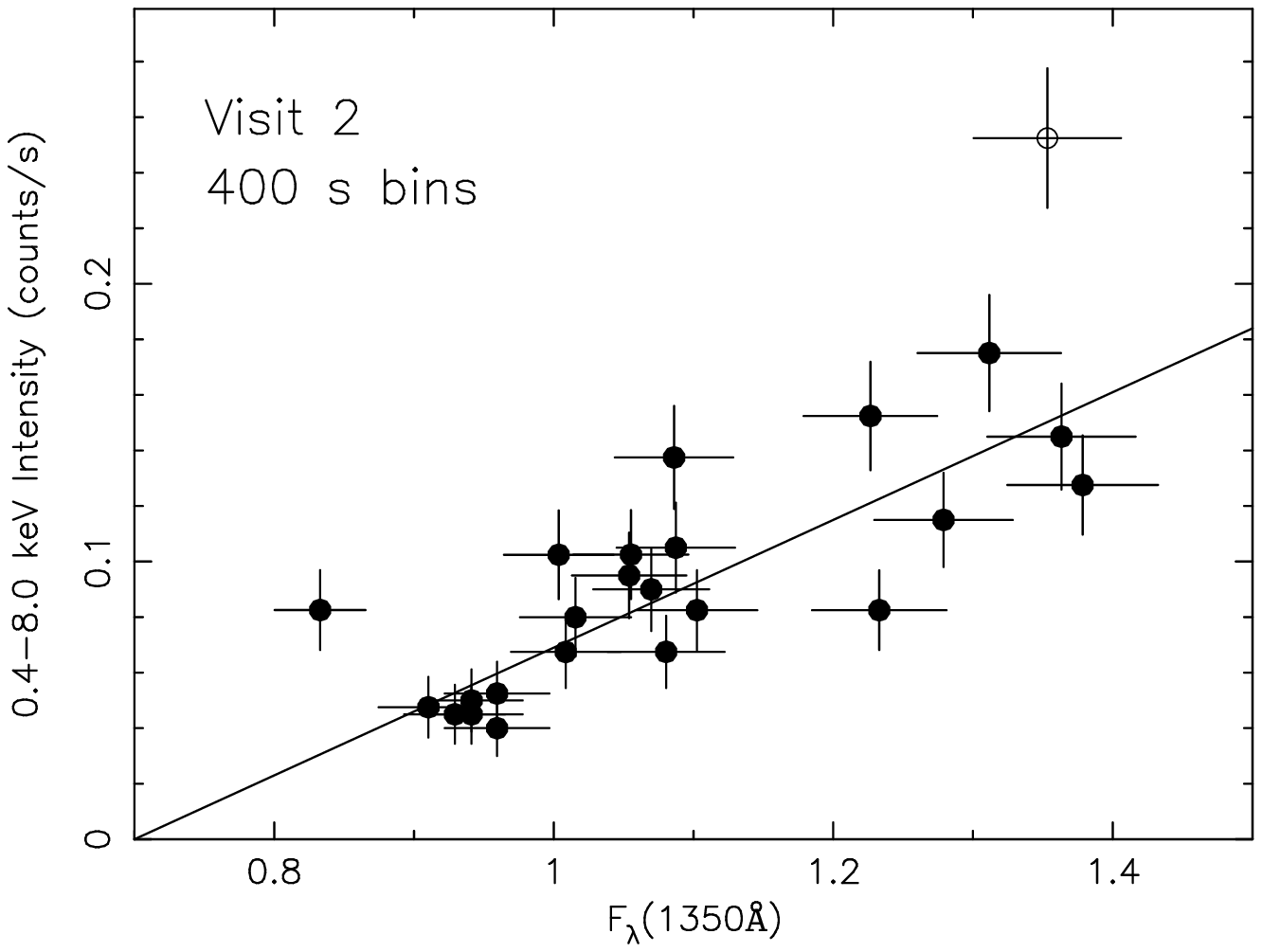}

\caption{Full-band  (0.4--8.0~keV)  intensity  versus  the UV  flux  at
1350~\r{A} during visit~2, using simultaneous 400~s time bins.  The UV
flux      in      both       panels      is      in      units      of
10$^{-15}$~erg~cm$^{-2}$~s$^{-1}$~\r{A}$^{-1}$.  The  line  shows  the
best-fitting  linear model,  which intercepts  the horizontal  axis at
$\sim$0.7. The datum indicated  with an open symbol was  not included in the
linear fit. \label{fig:visit2_uvxray_400}}

\end{figure}

We also  examined the  X-ray and UV  data that were  collected exactly
simultaneously.   The correlation  between  the X-ray  and UV  fluxes,
using    a    time    resolution     of    400~s,    is    shown    in
Fig.~\ref{fig:visit2_uvxray_400}.    As   shown   by   the   line   in
Fig.~\ref{fig:visit2_uvxray_400}, the  UV flux, relative  to the X-ray
intensity,        has       a       constant        component       of
$\sim\scinum{7}{-16}$~erg~cm$^{-2}$~s$^{-1}$~\r{A}$^{-1}$.          The
presence of this constant  component is noticeable also when comparing
the  variability of  the X-ray  and UV  light curves.   The normalized
excess variances of the simultaneous 0.4--8.0~keV and UV data shown in
Fig.~\ref{fig:visit2_uvxray_400}    are    $\scinuma{23}{4}{-2}$   and
$\scinuma{2.0}{0.3}{-2}$,  respectively.   Over   the  same  range  in
timescales the X-ray variability,  \emph{when normalized by the mean},
is thus clearly stronger than the UV variability.

It is  useful to  compare directly the  fluxes in  the \emph{variable}
components of the simultaneous X-ray  and UV data because the variable
UV  emission  is  possibly  the  result of  X-ray  reprocessing.   The
square-root  of  the \emph{unnormalized}  excess  variance (i.e.,  the
``excess       rms'')       in        the       UV       data       is
$\scinuma{1.5}{0.1}{-16}$~erg~cm$^{-2}$~s$^{-1}$~\r{A}$^{-1}$.     This
corresponds  to  a monochromatic  flux  variability  at 1350~\r{A}  of
$\scinum{2.0}{-13}$~erg~cm$^{-2}$~s$^{-1}$.   In the  X-ray  data, the
unnormalized              excess              variance              is
$\scinuma{4.7}{0.4}{-2}$~counts~s$^{-1}$.   To  convert this  counting
rate into a flux, we modeled the visit~2 X-ray spectrum over the range
1.3--8.0~keV using  a power law  modified by neutral  absorption.  The
flux     at    1~keV     of    the     underlying     power-law    was
$\scinum{5.0}{-13}$~erg~cm$^{-2}$~s$^{-1}$~keV$^{-1}$.   The  observed
variability in  the counting rate then corresponds  to a monochromatic
variability  at  1~keV of  $\scinum{2.4}{-13}$~erg~cm$^{-2}$~s$^{-1}$,
very similar to that of the UV.

Finally, we  note that  the mean monochromatic  flux at  1350~\r{A} is
$\scinum{1.4}{-12}$~erg~cm$^{-2}$~s$^{-1}$.   This   is  a  factor  of
$\sim$3  greater  than the  mean  monochromatic  flux  at 1~keV.   The
monochromatic    flux    of    the    constant   UV    component    is
$\scinum{9.3}{-13}$~erg~cm$^{-2}$~s$^{-1}$.

%%%%%%%%%%%%%%%%%%%%%%%%%%%%%%%%%%%%%%%%%%%%%%%%%%%%%%%%%%%%%%%%%%%%%%%%%%%%%
%%%%%%%%%%%%%%%%%%%%%%%%%%%%%%%%%%%%%%%%%%%%%%%%%%%%%%%%%%%%%%%%%%%%%%%%%%%%%

\section{DISCUSSION}

\subsection{Summary of Results}

We  have  analyzed  three  \emph{Chandra} observations  of  the  dwarf
Seyfert~1 NGC~4395.  The first  two observations, visit~1 and visit~2,
each have a  duration of $\sim$30~ks and are  separated by $\sim$50~ks
(end of visit 1 to beginning of visit 2), and were obtained as part of
a  multiwavelength  monitoring campaign.   The  third observation  was
obtained 4~years earlier  and has a duration of  $\sim$17~ks.  We have
also analyzed the  $\sim$90~ks \emph{XMM-Newton} observation from 2003
to facilitate a comparison with the \emph{Chandra} data.

The normalized excess variances  on timescales $\lesssim$15~ks show at
most  a weak  dependence  on energy  during  the three  \emph{Chandra}
observations,   while  the   \emph{XMM-Newton}   observation  exhibits
significantly softer variability.   An examination of the relationship
between  the   0.4--3.4~keV  and  3.4--8.0~keV   intensities  revealed
erratic,  short-term   (timescale  $\sim$1000~s)  spectral  variations
during visit~1,  which are seen  as deviations from the  mean relation
between the soft-band and  hard-band intensities.  During visit~2, the
relationship between the 0.4--2.7~keV and 4.2--8.0~keV intensities was
flatter        than       in       the        other       observations
(Fig.~\ref{fig:visit1_2_882_softvhard}).   During  observation~882 the
object reached higher intensities compared to visit~1 and visit~2.

Simultaneous UV observations were  conducted during part of the second
\emph{Chandra} observation. We detected a zero-lag correlation between
the  X-ray and  UV fluxes  with a  significance of  $\sim$99.5~\%. The
upper  limit  on any  lag  between the  two  light  curves is  roughly
$\pm$1~hour.

\subsection{X-ray Variability}

\subsubsection{Comparison with the XMM-Newton Observation}

The 2--10~keV fluxes during  visit~1, visit~2 and observation~882 were
$\sim\scinum{2.5}{-12}$,          $\sim\scinum{3.0}{-12}$          and
$\sim\scinum{3.5}{-12}$~erg~cm$^{-2}$~s$^{-1}$,    respectively.    In
comparison,   the   2--10~keV   flux  during   the   \emph{XMM-Newton}
observation  was  $\sim\scinum{5.5}{-12}$~erg~cm$^{-2}$~s$^{-1}$.  Our
variability   analysis    suggests,   then,   that    the   short-term
($\lesssim$15~ks) variability is significantly softer when the overall
flux level is higher.

\begin{figure}

\plotone{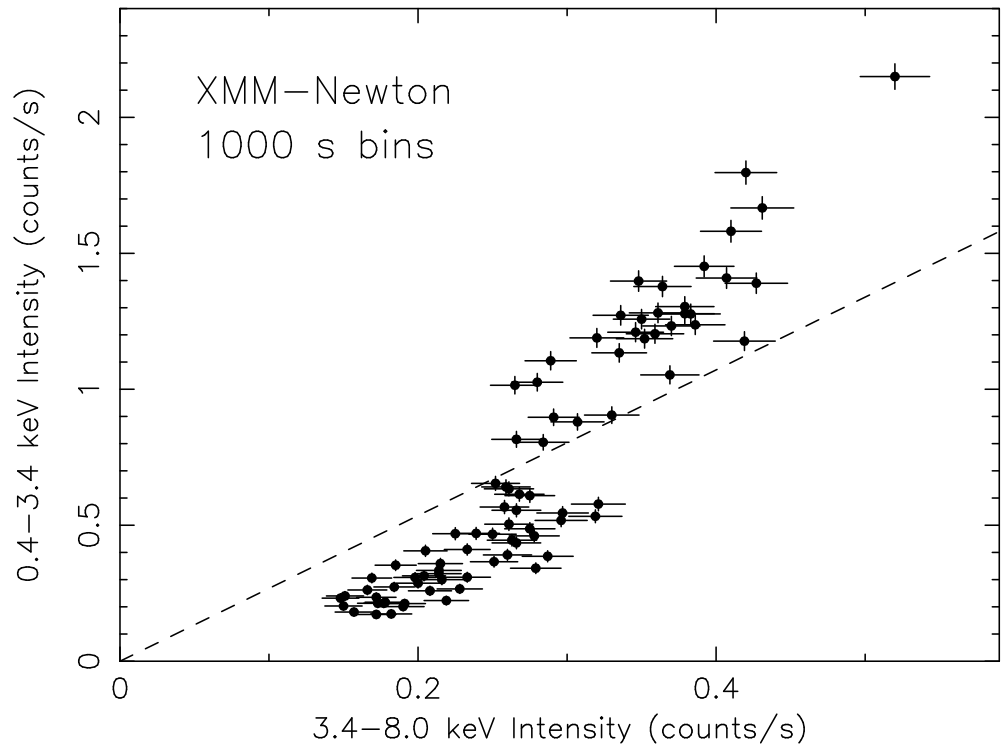}

\caption{Correlation   between  the  0.4--3.4~keV   and  3.4--8.0~keV
intensities  during  the  \emph{XMM-Newton}  observation  showing  the
softening  of   the  spectrum   with  increasing  flux.    Each  datum
corresponds to a time  bin of 1000~s.  The \emph{Chandra} observations
occupy roughly the region below the dashed line.\label{fig:xmm_softvhard}}

\end{figure}

In Fig.~\ref{fig:xmm_softvhard}  we show the  relationship between the
0.4--3.4~keV    and   3.4--8.0~keV    counting   rates    during   the
\emph{XMM-Newton} observation, with a time resolution of 1000~s. While
a detailed analysis of the  \emph{XMM-Newton} data is beyond the scope
of  this paper,  Fig.~\ref{fig:xmm_softvhard} shows  clearly  that the
softness increases  as the flux increases.   During the \emph{Chandra}
observations,  NGC~4395 occupied  only the  lower part  of  this track
(i.e., below the dashed line), with no excursions to the upper, softer
region (i.e., above  the dashed line).  (There is,  however, a hint in
Fig.~\ref{fig:visit1_2_882_softvhard}  that,  during  observation~882,
the  source  softness  increased  at  the time  of  highest  soft-band
counting  rate.)   NGC~4395   exhibited  stronger  flares  during  the
\emph{XMM-Newton}   observation   than   during   the   \emph{Chandra}
observations.  Consequently,  the \emph{XMM-Newton} data  exhibit both
higher flux and softer variability.

\subsubsection{Long-Term Spectral Variability}

The simplest model for explaining  the X-ray variability is to suppose
that variations  in a single  parameter, such as  mass-accretion rate,
produces  variability on  both  short and  long  timescales.  As  this
parameter  varies, the  object will  trace out  a track  in a  plot of
soft-band versus hard-band  intensity.  Different overall flux levels,
and spectral properties, then  correspond to different locations along
this track.  During visit~2, however,  a different track is traced out
compared  to  visit~1  and   observation~882.   This  shows  that  the
long-term    (i.e.,    \emph{inter}-observation,   $\gtrsim    30$~ks)
variability  between these two  observations is  in a  different sense
from the short-term (i.e.  \emph{intra}-observation, $\lesssim 30$~ks)
variability.    Between  visit~1  and   visit~2  the   mean  hard-band
(3.4--8.0~keV) counting  rate does not change  significantly while the
soft-band (0.4--3.4~keV) counting rate is scaled down.

One possibility for this spectral  change is that the underlying power
law may have  flattened between visit~1 and visit~2.  In the two-phase
model of  AGNs (and  also black hole  binary) X-ray emission,  the hot
phase (i.e.,  the corona)  inverse-Compton scatters seed  photons from
the cold phase (i.e., the disk) \citep[][]{sle76,st80,hm91,hm93}.  The
corona and disk  are radiatively coupled so that  some fraction of the
photons scattered in the corona return to the disk and are reprocessed
to  become seed  photons, which  for AGNs  are seen  presumably  at UV
wavelengths.   In this  scenario, the  photon index  of  the power-law
emission depends  on the rate at  which energy is  dissipated into the
corona, the physical properties of the corona, and the geometry of the
disk-corona system.

For a pair-dominated corona, variations in the compactness can produce
variations in the  photon index \citep[e.g.,][]{sps95,hmg97}.  In this
case,  a  decrease  in  the  luminosity  will  be  associated  with  a
flattening  of the  spectrum.  This  effect is  unlikely to  explain a
decrease  in  the  photon   index  between  the  visit~1  and  visit~2
\emph{Chandra} observations because  we observed no significant change
in the hard-band  counting rate. The compactness can  vary also if the
size  of the  emission region  varies. A  flattening of  the spectrum,
without a change in the source flux, could be caused by an increase in
the size of the corona.   Alternatively, for a corona not dominated by
pairs,  a variation  in optical  depth  can, in  principle, produce  a
change  in the  photon index  of  the spectrum  without an  associated
variation in the observed flux \citep[][]{hmg97}.

Variations in the photon index can  also be caused by a changes in the
accretion geometry,  as this  will influence the  amount by  which the
corona  is ``starved'' of  photons \citep[e.g.,][]{hmg94,sps95,zpm98}.
For  example, a  patchy  corona can  produce  relatively hard  spectra
because a reduction  in the coronal covering fraction  will reduce the
seed  flux that  enters the  corona.  Similarly,  a  centrally located
spherical corona  may be photon starved.  Variations  in geometry have
been  proposed to  explain  the spectral  variability associated  with
state changes  in galactic black  hole binaries \citep[][]{emn97,z98}.
These seem highly unlikely in the case of NGC 4395, however, given the
lack  of  hard  flux  variability  and  the  short  timescale  of  the
variations,  which correspond  only to  a  few seconds  in a  galactic
binary.  State changes are usually  seen on much longer time scales (a
few  days),   although  an  exception  to  this   is  the  microquasar
GRS~1915$+$105.   In  this   object,  unusual  ``state''  changes  are
observed on a timescale  of seconds \citep[][]{bmk97b,bmk97a}, but are
accompanied by large changes in flux, so cannot apply in NGC~4395.

The fundamental  problem with intrinsic  continuum changes is  that to
get  a  large change  in  spectral index  without  any  change in  the
hard-band counting  rate requires  a high degree  of fine  tuning. Any
pivoting of the  spectrum would need to be  counteracted exactly by an
appropriate variation in the flux normalization, which would depend on
the pivot energy.  Given this,  arguably a more likely explanation for
the  scaling  down of  the  soft-band  intensity  between visit~1  and
visit~2   is  an  increase   in  absorption.    \citet[][]{ifa00}  and
\citet[][]{sif03} have already shown  evidence of a multi-zone ionized
absorber in this object, and attributed short-term spectral changes to
variations of  the ionization parameter  in response to  the continuum
flux.  We  can rule out this interpretation  for the inter-observation
spectral variability between visits~1 and 2 because the mean hard-band
intensity  does not  change between  the observations.   Rather, there
would need to be an increase in the column density of the absorber, or
a change  in ionization unrelated  to the continuum.   Recent detailed
modeling of  UV and X-ray absorbers  of some sources  have shown clear
evidence for variations  of absorbing gas due to  transverse motion of
gas through  the line of sight  \citep[e.g.,][]{gck03,kgc05}, and this
appears to be the most  likely mechanism to account for the variations
in NGC~4395.

\subsubsection{Short-Term Spectral Variability}

Significant scatter about the mean relation between the counting rates
in two bands has been  seen previously in Seyfert galaxies \citep[see,
e.g., the ``flux-flux plots'' of][]{tum03}. NGC~4395 exhibited erratic
spectral variations during visit~1,  and these were seen as deviations
from the mean relation  between the 0.4--3.4 and 3.4--8.0~keV counting
rates.    Unfortunately,  our   data   do  not   have  a   high-enough
signal-to-noise ratio to investigate the nature of these variations in
detail. We  consider below possible explanations for  how the spectral
variations might be  produced either within, or outside  of, the X-ray
emission region.

One explanation  was mentioned by \citet[][]{mel05} in  the context of
observation~882.   Those authors found  significant variations  in the
ratio between the counting  rates in 0.3--1.2~keV and 2--10~keV (i.e.,
different from  our ``soft'' and  ``hard'' bands), and  they suggested
that these variations are produced by changes in the column density of
a single-zone  warm absorber.  In  this kind of scenario,  the erratic
variations are superimposed on the existing short-term variability.

An alternative  possibility is that  the \emph{underlying} variability
in the  various energy  bands becomes uncorrelated  at times.   At the
most  basic level, the  loss of  coherence between  the soft  and hard
intensities rules out  the simplest, single-temperature Comptonization
models.   In these models,  any spectral  variations should  be smooth
\citep[e.g.,][]{hmg97}.

The presence of scatter about  the mean relation between the soft-band
and  hard-band  counting  rates  is  possibly  a  manifestation  of  a
hardening of the  power spectrum at high frequencies,  as is seen, for
example,    in    NGC~7469    \citep[][]{np01}    and    MCG$-$6-30-15
\citep[][]{vfn03,vf04,pka05}.    In  the   context   of  ``propagating
perturbation''  models \citep[][]{l97,cgr01,kcg01}, variations  in the
mass accretion rate  are injected into the accretion  flow over a wide
range of disk radii, with the timescale of the variations being longer
for larger  radii. These perturbations then propagate  inward and give
rise  to $f^{-1}$  variations in  the accretion  rate at  the emission
region. The variations at  frequencies greater than the power-spectral
break are assumed to originate from within the emission region itself.
The inner  parts of this  region experience faster  perturbations, and
produce harder spectra,  than the outer parts, thus  causing the power
spectrum  above  the  break   frequency  to  flatten  with  increasing
energy. In  NGC~4395, this power-spectral  break is at a  timescale of
$\sim$500~s  \citep[][see  \S~1]{vif04}.   Our  correlation  analysis,
however, does not probe these short timescales, so we cannot appeal to
propagating perturbations to explain the observed lack of coherence.

One alternative is to invoke a shot-type model \citep[e.g.,][]{t72} in
which there  are many  short-lived, disconnected emitting  regions, or
`flares'. If these regions possess differing spectral properties, then
the spectral shape of their  combined X-ray emission might be expected
to vary.  Note, however, that  shot-noise models are challenged by the
combined presence  of a linear  rms-flux correlation and  a log-normal
distribution of fluxes \citep[][]{um01,u04,umv05}.

\subsection{X-ray--UV Connection}

The two-phase  model \citep[][]{hm91} predicts  that the X-ray  and UV
emission  should be  linked \citep[see  also][]{gr88}.  The  number of
objects  in  which a  definite  link  has  been observed  is  somewhat
controversial  \citep[see][and references  therein]{mme02}.   The best
examples   thus   far   are   NGC~5548   \citep[][]{cnm92},   NGC~4151
\citep[][]{eac96}, and  NGC~7469 \citep[][]{nlg00,pmh04}. In  the last
object, a  correlation was seen between  the UV flux  and X-ray photon
index, rather  than between the  UV and X-ray  fluxes. In a  number of
other  objects there  is a  very poor  correlation between  the bands,
particularly on short time scales \citep[e.g.,][]{dwf90,ekn00}.

The  zero-lag  correlation between  the  X-ray  and  UV fluxes  during
visit~2  has a  detection significance  of $\sim$99.5~\%.   This  is a
conservative estimate as it accounts for the possibility of the chance
alignment  of two  independent flicker-noise  light  curves.  NGC~4395
thus satisfies a basic prediction of two-phase models.  The X-ray flux
illuminates  the  material  responsible  for the  (pseudo-thermal)  UV
emission (e.g.,  the accretion disk),  which in turn  reprocesses them
into UV seed photons to be Compton upscattered into the X-ray band.

The two-phase  model predicts that  the luminosity of  the reprocessed
component  should  have the  same  order  of  magnitude as  the  X-ray
luminosity \citep[][]{hmg94}.  This implies  that the amplitude of the
UV flux variability should be roughly  the same as that of the X-rays.
We determined  the fluxes of the  variable X-ray and  UV components by
measuring the  square-root of the  \emph{unnormalized} excess variance
of the X-ray  and UV data that were  collected exactly simultaneously.
The      monochromatic       flux      variability      was      $\sim
\scinum{2}{-13}$~erg~cm$^{-2}$~s$^{-1}$ at  both 1350~\r{A} and 1~keV.
We must stress that the  comparison we perform here is very uncertain;
a rigorous analysis  would require broad-band luminosity measurements.
This is,  unfortunately, hampered by the difficulties  in modeling the
ionized absorber in  the X-ray spectrum and the  fact that useful data
cannot be obtained in the extreme UV band. Nevertheless, the fact that
the X-ray and  UV flux variabilities have the  same order of magnitude
suggests that  an interpretation within the  two-phase model framework
is energetically reasonable.

The   relationship  between  the   X-ray  and   UV  fluxes   shown  in
Fig.~\ref{fig:visit2_uvxray_400} suggests that  the UV flux contains a
non-reprocessed             component             of             $\sim
\scinum{7}{-16}$~erg~cm$^{-2}$~s$^{-1}$~\r{A}$^{-1}$, which is roughly
65~\% of the  mean flux.  In the context of  the two-phase model, this
flux may be attributed to  intrinsic emission from the accretion disk.
In this scenario, only a  portion of the accretion power is dissipated
in the corona, which itself must be patchy \citep[][]{hmg94,hmg97}.

From a  cross-correlation analysis,  the upper limit  on the  time lag
between the  X-ray and UV  fluxes is about $\pm$1~hour.   The distance
between the  reprocessor and the corona must,  therefore, be separated
by a distance  not greater than $\sim$10$^{14}$~cm.  For  a black hole
mass of $\mbh =  \scinuma{3.6}{0.6}{5}$~$\msun$, this corresponds to a
distance  of about  2000~$r_{\mathrm{g}}$ ($r_{\mathrm{g}}  =  G\mbh /
c^{2}$). The  upper limit  on the time  lag we measured  is consistent
with the measured  lag between the UV continuum  and line fluxes found
by \citet{pbd05}.

It  is  worth noting  here  that  \citet[][see  also Desroches  et~al.
2006]{slp05} observed $\sim$10~\%  variations in the optical continuum
of NGC~4395,  over timescales of roughly 30~minutes  to 8~hours.  They
pointed  out that  this optical  variability is  much more  rapid than
expected  if the  variability is  the result  of instabilities  in the
disk.  The rapid variations are, however, quite naturally explained in
the context  of X-ray  reprocessing.  The current  puzzle is  why this
mechanism appears to operate  in some sources, including NGC~4395, but
not  in  others.  A  possible  solution can  be  seen  in the  complex
behavior we see in  NGC~4395, which suggests several different origins
for  the flux  and spectral  variability (e.g.   in the  seed photons,
intrinsic  to  corona,   associated  with  absorption).   Clearly  the
presence or  absence of  any UV/X-ray correlation  will depend  on the
dominant mode of variability at the time of a given observation.

\subsection{A Real Periodicity at 341~s?} 

While many claims have been made, no periodic or quasi-periodic signal
has  yet been  robustly detected  in an  AGN  \citep[see][]{v05}.  The
detection significance  of 99.0~\% that we calculated  for the feature
at  341~s in the  visit~1 power  spectrum represents  one of  the best
candidate signals found thus far.  A feature at 390~s was present also
during  the   second  half   of  observation~882,  with   a  detection
significance   of  95.0~\%,   in  agreement   with  the   analysis  of
\citet[][]{mel05}.

This  power-spectral  feature,  however,  should be  interpreted  very
cautiously. First,  it was not  detected during either visit~2  or the
\emph{XMM-Newton}  observation.  Having  searched for  the  feature in
these data,  the overall ``number-of-trials''  is thus more  than just
the number  of independent frequences  in the visit~1  power spectrum.
Second, excess power  was found only in a  single frequency bin, which
would correspond to a quality  factor ($Q = \nu/FWHM$) of greater than
about 90.  In contrast, the \emph{quasi}-periodic oscillations seen in
galactic  black hole  binaries  are broader  than  this, with  quality
factors  of up  to only  about 30  \citep[][]{rmm02}.  Given  that the
broad-band variability is similar in  AGNs and black hole binaries, we
might   also  expect  similarities   with  regard   to  quasi-periodic
oscillations.   A genuine  detection cannot  yet be  claimed, although
NGC~4395 remains a promising candidate for future studies.

%%%%%%%%%%%%%%%%%%%%%%%%%%%%%%%%%%%%%%%%%%%%%%%%%%%%%%%%%%%%%%%%%%%%%%%%%%%%%
%%%%%%%%%%%%%%%%%%%%%%%%%%%%%%%%%%%%%%%%%%%%%%%%%%%%%%%%%%%%%%%%%%%%%%%%%%%%%

\acknowledgments

We  thank Harvey  Tananbaum, the  Chandra X-ray  Center  Director, for
awarding  Director's  Discretionary  Time.   We  thank  also the  anonymous
referee for  constructive comments  that improved the  manuscript. PMO
acknowledges financial support from PPARC.  AL acknowledges support by
the Israel Science  Foundation (Grant \#1030/04), and by  a grant from
the Norman and Helen Asher  Space Research Institute.  SK is supported
in  part at  the  Technion by  a  Zeff Fellowship.   This research  is
supported  by NASA  through {\it  HST} grant  GO-09818 from  the Space
Telescope Science  Institute, which is operated by  the Association of
Universities  for Research  in  Astronomy, Inc.,  under NASA  Contract
NAS5-26555.  A.V.F.   is also grateful for the  assistance of National
Science Foundation (NSF) grant  AST-0307894.  L.-B.D.  is supported by
a  Julie-Payette/Doctoral  Fellowship from  the  Natural Sciences  and
Engineering  Research Council  of  Canada and  by  the Canadian  Space
Agency.

%%%%%%%%%%%%%%%%%%%%%%%%%%%%%%%%%%%%%%%%%%%%%%%%%%%%%%%%%%%%%%%%%%%%%%%%%%%%%
%%%%%%%%%%%%%%%%%%%%%%%%%%%%%%%%%%%%%%%%%%%%%%%%%%%%%%%%%%%%%%%%%%%%%%%%%%%%%

%%%%%%%%%%%%%%%%%%%%%%%%%%%%%%%%%%%%%%%%%%%%%%%%%%%%%%%%%%%%%%%%%%%%%%%%%%%%%
%%%%%%%%%%%%%%%%%%%%%%%%%%%%%%%%%%%%%%%%%%%%%%%%%%%%%%%%%%%%%%%%%%%%%%%%%%%%%

%%%%%%%%%%%%%%%%%%%%%%%%%%%%%%%%%%%%%%%%%%%%%%%%%%%%%%%%%%%%%%%%%%%%%%%%%%%

%%%%%%%%%%%%%%%%%%%%%%%%%%%%%%%%%%%%%%%%%%%%%%%%%%%%%%%%%%%%%%%%%%%%%%%%%%%

%%%%%%%%%%%%%%%%%%%%%%%%%%%%%%%%%%%%%%%%%%%%%%%%%%%%%%%%%%%%%%%%%%%%%%%%%%%

%%%%%%%%%%%%%%%%%%%%%%%%%%%%%%%%%%%%%%%%%%%%%%%%%%%%%%%%%%%%%%%%%%%%%%%%%%%

%%%%%%%%%%%%%%%%%%%%%%%%%%%%%%%%%%%%%%%%%%%%%%%%%%%%%%%%%%%%%%%%%%%%%%%%%%%

%%%%%%%%%%%%%%%%%%%%%%%%%%%%%%%%%%%%%%%%%%%%%%%%%%%%%%%%%%%%%%%%%%%%%%%%%%%

%%%%%%%%%%%%%%%%%%%%%%%%%%%%%%%%%%%%%%%%%%%%%%%%%%%%%%%%%%%%%%%%%%%%%%%%%%%

%%%%%%%%%%%%%%%%%%%%%%%%%%%%%%%%%%%%%%%%%%%%%%%%%%%%%%%%%%%%%%%%%%%%%%%%%%%

%%%%%%%%%%%%%%%%%%%%%%%%%%%%%%%%%%%%%%%%%%%%%%%%%%%%%%%%%%%%%%%%%%%%%%%%%%%

\clearpage

%%%%%%%%%%%%%%%%%%%%%%%%%%%%%%%%%%%%%%%%%%%%%%%%%%%%%%%%%%%%%%%%%%%%%%%%%%%%%
%%%%%%%%%%%%%%%%%%%%%%%%%%%%%%%%%%%%%%%%%%%%%%%%%%%%%%%%%%%%%%%%%%%%%%%%%%%%%

\begin{deluxetable}{lccccc}

\tablecaption{Normalized excess variance measurments. \label{tab:snxs}}

\tablehead{

\colhead{Energy   Range}   &   \multicolumn{5}{c}{Normalized   Excess   Variance
($\times$10$^{-2}$)} \\

\colhead{(keV)}  &  \colhead{Visit 1}  &  \colhead{Visit 2}  &  \colhead{Obs. 882}  &
\colhead{\emph{Chandra}} & \colhead{\emph{XMM-Newton}} 

}

\startdata

0.4--8.0    &   $\decnuma{10.7}{1.4}$   &    $\decnuma{10.6}{1.6}$   &
$\decnuma{16.2}{1.9}$ &  $\decnuma{11.8}{0.9}$ & $\decnuma{23.7}{0.4}$
\\

0.4--3.4    &   $\decnuma{11.0}{2.0}$   &    $\decnuma{15.5}{3.3}$   &
$\decnuma{19.9}{3.1}$ &  $\decnuma{14.5}{1.7}$ & $\decnuma{33.0}{0.6}$
\\

3.4--8.0    &   $\decnuma{10.9}{2.4}$    &    \phn$\decnuma{7.4}{1.9}$   &
$\decnuma{12.6}{2.6}$ & \phn$\decnuma{9.8}{1.3}$ & \phn$\decnuma{7.9}{0.4}$ \\

0.4--2.7    &   $\decnuma{13.5}{2.8}$   &    $\decnuma{11.8}{3.9}$   &
$\decnuma{21.5}{4.0}$ &  $\decnuma{14.4}{2.1}$ & $\decnuma{36.3}{0.6}$
\\

2.7--4.2    &   \phn$\decnuma{9.4}{2.7}$    &    $\decnuma{11.3}{3.2}$   &
$\decnuma{19.7}{4.3}$ &  $\decnuma{12.2}{1.9}$ & $\decnuma{11.4}{0.7}$
\\

4.2--8.0    &    \phn$\decnuma{9.1}{2.9}$    &   \phn$\decnuma{7.9}{2.5}$    &
\phn$\decnuma{8.8}{3.0}$ & \phn$\decnuma{8.6}{1.6}$ & \phn$\decnuma{7.0}{0.5}$ \\

\enddata
\tablecomments{~Light-curve duration 15~ks, time resolution 600~s.}
\end{deluxetable}

%%%%%%%%%%%%%%%%%%%%%%%%%%%%%%%%%%%%%%%%%%%%%%%%%%%%%%%%%%%%%%%%%%%%%%%%%%%%%
%%%%%%%%%%%%%%%%%%%%%%%%%%%%%%%%%%%%%%%%%%%%%%%%%%%%%%%%%%%%%%%%%%%%%%%%%%%%%

\end{document}